\documentclass[12pt]{article}
\usepackage{epsf}
\usepackage{amsmath}
\usepackage{graphics}
\usepackage{cite}

\renewcommand{\thefootnote}{\#\arabic{footnote}}
\begin{document}

\newcommand{\gtrsim}{ \mathop{}_{\textstyle \sim}^{\textstyle >} }
\newcommand{\lesssim}{ \mathop{}_{\textstyle \sim}^{\textstyle <} }

\newcommand{\rem}[1]{{\bf #1}}

\renewcommand{\thefootnote}{\fnsymbol{footnote}}
\setcounter{footnote}{0}
\begin{titlepage}

\def\thefootnote{\fnsymbol{footnote}}

\begin{center}
\hfill March 2016\\
\vskip .5in
\bigskip
\bigskip
{\Large \bf Searching for Dark Matter Constituents with Many Solar Masses}

\vskip .45in

{\bf Paul H. Frampton\footnote{email: paul.h.frampton@gmail.com\\
homepage:www.paulframpton.org}}

{\em 4 Lucerne Road, Oxford OX2 7QB, UK.}

\end{center}

\vskip .4in
\begin{abstract}
\noindent
Searches for dark matter (DM) constituents are presently mainly focused on axions and
WIMPs despite the fact that far higher mass constituents are viable. 
We discuss and dispute whether axions exist and those arguments for WIMPs
which arise from weak scale supersymmetry. We focus on the highest possible
masses and argue that, since if they constitute all DM they cannot be baryonic,
they must uniquely be primordial black holes. Observational constraints require them
to be of intermediate masses mostly between a hundred
and a hundred thousand solar masses. Known search strategies for such PIMBHs
include wide binaries, CMB distortion
and, most promisingly, extended microlensing experiments.
\end{abstract}

\end{titlepage}

\renewcommand{\thepage}{\arabic{page}}
\setcounter{page}{1}
\renewcommand{\thefootnote}{\#\arabic{footnote}}

\newpage

\section{Introduction}

\noindent
Astronomical observations have led to a consensus that the energy
make-up of the visible universe is approximately 70\% dark energy, 25\% dark matter
and only 5\% normal matter. The dark energy remains mainly mysterious; the 
dark matter problem will be addressed in the present paper; and the
normal matter has a successful theory applicable up to at least a few hundred GeV 
in the form of the standard model\footnote{More precise
numbers from the Planck Collaboration\cite{Planck} are 68.4\% dark energy, 26.7\% dark matter
and 4.9\% normal matter.}.

\bigskip

\noindent
General discussions of the history and experiments for dark matter are in
\cite{Sciama,Sanders,Bertone}. A recent popular book
\cite{Freese} is strong on the panoply of unsuccessful WIMP searches.

\bigskip

\noindent
In the present article we shall make the unjustified assumption that there is only
one species of dark matter. Because the luminous
matter is far richer than this, with its varied menu of the three families of quarks and leptons, 
the gauge bosons
and the BEH boson, there is no sharp reason why the dark matter should be
so different and simpler. One practical reason to make such an assumption
is to simplify the research. A better reason is that it is likely, in our opinion, to be correct.
If it is not so, then our conclusions may apply only to one component, perhaps
the dominant component of the dark matter.

\bigskip

\noindent
The present ignorance of the dark matter sector is put into perspective
by looking at the uncertainty in the values of the constituent
mass previously considered. The lightest such candidate is the invisible axion
with $M = 1\mu eV$. One very massive such candidate is
the intermediate mass
black hole (IMBH) with $M = 100,000 M_{\odot}$ which is
a staggering seventy-seven orders of magnitude larger. Our aim
is to reduce this uncertainty.

\bigskip

\noindent
A result of the present analysis will be that the number of
orders of magnitude uncertainty in the dark matter constituent
mass can be reduced to three. We shall conclude,
after extensive discussion, that the most viable candidate for the constituent
which dominates dark matter is
the intermediate mass black hole (IMBH) with mass in the range
\begin{equation}
100 M_{\odot} < M_{IMBH} < 100,000 M_{\odot}
\label{IMBH}
\end{equation}

\bigskip

\noindent
Less experimental effort is being
invested in searching for IMBHs than for WIMPs. WIMP
searches include 
terrestrial direct detection, astronomical indirect detection and 
production of WIMPs at the LHC.

\bigskip

\noindent
One reason for the neglect of IMBHs may be that the literature is confusing
including one study which claimed entirely to rule out Eq.(\ref{IMBH}). We shall attempt
to clarify the situation which actually still permits the whole range in Eq.(\ref{IMBH}).
The present paper is, in part, an
attempt to redress the imbalance between the few
experimental efforts to search for IMBHs
compared to the extensive WIMP searches. 

\bigskip

\noindent
One possible reason for previously overlooking
our solution to the dark matter problem is that it had been assumed that all black holes
arise only from gravitational collapse of baryonic objects, either normal stars
or superheavy early stars. 

\bigskip

\noindent
Some of this paper is review of old topics but it is with an original viewpoint to
try to identify what is the most likely constituent of the dark matter, more by logic
and partial elimination of popular possibilities than by new calculations.

\bigskip

\noindent
The plan of this paper is that in Section 2 we discuss 
both axions and WIMPS then
in Section 3 examples of MACHOs including the IBMHs are discussed. Three experimental
methods to search for IMBHs are discussed in Sections 4 (wide binaries),
Section 5 (distortion of the CMB) and Section 6 (microlensing). 
Finally Section 7 is devoted to further discussion.

\bigskip
\bigskip
\bigskip

\section{Axions and WIMPs}

\bigskip

\subsection{Axions}

\noindent
It is worth reviewing briefly the history of the axion particle now believed,
if it exists, to lie in the mass range
\begin{equation}
10^{-6} eV < M < 10^{-3} eV
\label{axion}
\end{equation}

\noindent
The lagrangian originally proposed for Quantum Chromodymamics (QCD) was of the
simple form, analogous to Quantum Electrodynamics,
\begin{equation}
{\cal L}_{QCD} = -\frac{1}{4} G_{\mu\nu}^{\alpha}G^{\mu\nu}_{\alpha}
- \frac{1}{2}  \sum_i \bar{q}_{i,a}\gamma^{\mu}D^{ab}_{\mu} q_{i,b}
\label{QCD}
\end{equation}
summed over the six quark flavors.

\bigskip

\noindent
The simplicity of Eq.(\ref{QCD}) was only temporary and became more
complicated
in 1975 by the discovery of instantons
\cite{Belavin} which dictated\cite{Hooft1,Hooft2} that an additional term in the QCD lagrangian
must be added
\begin{equation}
\Delta {\cal L}_{QCD} = \frac{\Theta}{64\pi^2} G_{\mu\nu}^{\alpha} \tilde{G}^{\mu\nu}_{\alpha}
\label{GGdual}
\end{equation}
where $\tilde{G}_{\mu\nu}$ is the dual of $G_{\mu\nu}$. Although this extra term is an exact derivative, it cannot be discarded as a surface term because
there is now a topologically nontrivial QCD vacuum with an infinite number of different
values of the spacetime integral over Eq.(\ref{GGdual}) all of which correspond to $G_{\mu\nu}^{\alpha}=0$. Normalized as in Eq.(\ref{GGdual}), the spacetime integral
of this term must be an integer, and an instanton configuration
changes this integer, or Pontryagin number, by unity.

\bigskip

\noindent
When the quark masses are complex, an instanton changes not only $\Theta$ but also
the phase of the quark mass matrix ${\cal M}_{quark}$ and the full phase to be considered
is
\begin{equation}
\bar{\Theta} = \Theta + \arg \det ||{\cal M}_{quark}||
\label{thetabar}
\end{equation}

\bigskip

\noindent
The additional term, Eq.(\ref{GGdual}),violates P and CP, and contributes to the neutron
electric dipole moment whose upper limit \cite{Baker} provides a constraint
\begin{equation}
\bar{\Theta} < 10^{-9}
\label{strongCP}
\end{equation}
which fine-tuning is the strong CP problem.

\bigskip

\noindent
The hypothetical axion particle then arises from an ingenious technique to resolve
Eq.(\ref{strongCP}), although as it turns out it may have been too ingenious. The
technique is based on the Peccei-Quinn mechanism\cite{PQ1,PQ2} which introduces a new
global
$U(1)_{PQ}$ symmetry which allows the vacuum to relax to $\bar{\Theta} = 0$. Because
this $U(1)_{PQ}$ symmetry is spontaneously broken, it gives rise \cite{Weinberg,Wilczek}
to a light pseudoscalar axion with mass in the range $100keV < M < 1 MeV$.
An axion in this mass range was excluded experimentally but then the theory
was modified to one with an invisible axion\cite{Dine2,Kim,Zhit,SVZ} where the
$U(1)_{PQ}$ symmetry is broken at a much higher scale $f_a$ and the coupling
of the axion correspondingly suppressed. Nevertheless, clever experiments to detect
such so-called invisible axions were proposed\cite{Sikivie1}.

\bigskip

\noindent
Over twenty years ago, in 1992, three papers\cite{Holman,Kamionkowski,BarrSeckel}
independently pointed out a serious objection to the invisible axion. The point is that the 
invisible axion
potential is so fine-tuned that adding gravitational couplings for
weak gravitational fields at the dimension-five level requires tuning of a dimensionless
coupling $g$ to be at least as small as
$g < 10^{-40}$, more extreme than the tuning of $\bar{\Theta}$
in Eq.(\ref{strongCP}) that one is trying to avoid.

\bigskip

\noindent
Although a true statement, it is not a way out of this objection
to say that we do not know the correct theory of quantum gravity
because for weak gravitational fields, as is the case almost everywhere
in the visible universe, one can use an effective field theory 
as discussed in  \cite{Donoghue}. To our knowledge, this serious objection 
to the invisible axion which has been generally ignored since 1992 has not gone away and
therefore the invisible axion may not exist. This issue is not settled but in a study
of dark matter, we may {\it pro tempore} assume that there is no axion.

\bigskip

\noindent
There would remain the strong CP problem of Eq.(\ref{strongCP}). One other solution would be
a massless up quark but this is disfavored by lattice calculations\cite{Nelson2}.
For the moment, Eq.(\ref{strongCP}) must be regarded as fine tuning. We recall that the
ratio of any neutrino mass to the top quark mass in the standard model satisfies  
\begin{equation}
\left( \frac{M_{\nu}}{ M_t } \right) < 10^{-12}. 
\label{nut}
\end{equation}

\subsection{WIMPs}

\noindent
By Weakly Interacting Massive Particle (WIMP) is generally meant an unidentified elementary
particle with mass in the range, say, between 10 GeV and 1000 GeV and with scattering cross
section with nucleons ($N$) satisfying, according to the latest unsuccessful WIMP direct searches, 
\begin{equation}
\sigma_{WIMP-N} < 10^{-44} cm^2
\label{WIMPcc}
\end{equation}
which is somewhat smaller than, but roughly comparable to, the characteristic strength of the known weak interaction.

\bigskip

\noindent
The WIMP particle must be electrically neutral and be stable or have an extremely
long lifetime. In model-building, the stability may be achieved by an {\it ad hoc} discrete
symmetry, for example a $Z_2$ symmetry under which all the standard model
particles are even and others are odd. If the discrete symmetry is unbroken,
the lightest odd state must be stable
and therefore a candidate for a dark matter. In general, this appears contrived because
the discrete symmetry is not otherwise motivated.

\bigskip

\noindent
By far the most popular WIMP example came from weak scale supersymmetry where a
discrete R symmetry has the value R=+1 for the standard model particles
and R=-1 for all the sparticles. Such an R parity is less {\it ad hoc} 
being essential to prevent too-fast
proton decay. The lightest R=-1 particle is stable and, if
not a gravitino which has the problem of too-slow decay in the early universe,
it was the neutralino\cite{Goldberg}, a linear combination of zino, bino and
higgsino. The neutralino provided an attractive candidate.

\bigskip

\noindent
A problem for the neutralino is at the LHC 
where weak scale supersymmetry not many years ago confidently predicted
sparticles (gluinos, etc.) at the weak scale $\sim 250$ GeV
there is no sign of a gluino with mass anywhere up to at least 1700 GeV
\cite{ATLASsusy,CMSsusy,CMS2} so weak scale supersymmmetry
may not exist. Nevertheless the jury is still out and and numerous physicists
remain optimistic. For present purposes we can, again {\it pro tempore},
assume there is no WIMP. 

\bigskip

\noindent
It is worth briefly recalling the history of weak scale supersymmetry. The standard
model\cite{SLG,Wein,GIM,GerardHooft} was in place by 1971 and its biggest
theoretical problem was that, unlike QED with only log divergences, the scalar
sector of the standard model generates quadratic divergences which 
destabilize the mass of the BEH boson.

\bigskip

\noindent
When supersymmetric field theories were invented\cite{WessZumino} in 1974,
they provided an elegant solution of the quadratic divergence problem
and hence immediately became popular. Even more so in 1983 when
the neutralino was identified\cite{Goldberg} as a dark matter candidate and more so again
in 1991 when it was pointed out\cite{ADF} that grand unification works better with 
the supersymmetric partners included.

\bigskip

\noindent
With the benefit of hindsight, these motivations for supersymmetry can all be otherwise realized. The quadratic divergence can cancel\cite{CFR}  in non-supersymmetric quiver theories. 
A dark matter candidate can be invented, in an {\it ad hoc} fashion, within
conformality model building\cite{DMconformality}. Historically, the neutralino appeared in particle phenomenology research {\it before} the WIMP acronym entered\cite{KT} the lexicon of cosmology. It is an important point that the WIMP idea came from weak scale supersymmetry. 

\bigskip

\noindent
Precise unification\cite{ADF} with supersymmetry by adding one parameter, a common
sparticle mass, was not miraculous but had at least a 20\% probability as shown in \cite{ADFFL}. Other precise
grand unifications \cite{ 4TeV} are known without supersymmetry
in conformality model building. 
If we do need\footnote{The unnaturalnesses exhibited in
Eq.(\ref{strongCP}) and Eq.(\ref{nut}) of the text both may suggest new physics beyond the standard model.} 
As a replacement for weak scale supersymmetry, 
conformal invariance is a contender as discusssed in 1998\cite{conf1}
and a number of subsequent papers 
as well more recently in \cite{Hooft} and\cite {Mannheim}.

\bigskip

\noindent
Run 2 of the LHC is not necessarily doomed if WIMPs and sparticles do not exist.
An important question, independent of naturalness but surely related to anomalies,
is the understanding of why there are three families of quarks and leptons. For that reason
Run2 may discover additional gauge bosons, siblings of the $W^{\pm}$ and
$Z^0$, as discussed in \cite{PHF331,331}.

\bigskip

\noindent
The many
attempts to detect WIMPs directly and indirectly are discussed in \cite{Freese}.

\bigskip
\bigskip
\bigskip

\section{MACHOs}

\noindent
Massive Compact Halo Objects (MACHOs) are commonly defined\cite{Griest}
by the notion of compact objects used in astrophysics\cite{Shapiro} as
the end products of stellar evolution when most of the nuclear fuel has been expended.
They are usually defined to include white dwarfs, neutron stars, black holes, brown dwarfs
and unassociated planets, all equally hard to detect because they do not emit any
radiation.

\bigskip

\noindent
This narrow definition implies, however, that MACHOs are composed of normal matter
which is too restrictive in the special case of black holes.
It is here posited that black holes of arbitrarily
high mass up to $100,000 M_{\odot}$ can be
produced primordially as calculated and demonstrated in \cite{Yanagida}.
Nevertheless the acronym MACHO still nicely
applies to dark matter IMBHs which are
massive, compact, and in the halo.

\bigskip

\noindent
Unlike the axion and WIMP elementary particles which would have a definite mass, the
black holes will have a range of masses. The lightest PBH which has
survived for the age of the universe has a lower mass limit
\begin{equation}
M_{PBH} > 10^{-18} M_{\odot} \sim 10^{36} TeV
\label{PBHmin}
\end{equation}
already thirty-six orders of magnitude heavier than the heaviest would-be WIMP.
This lower limit comes from the lifetime formula derivable from Hawking radiation
\cite{Hawking}
\begin{equation}
\tau_{BH}(M_{BH}) \sim \frac{G^2 M_{BH}^3}{\hbar c^4} 
\sim 10^{64} \left( \frac{M_{BH}}{M_{\odot}} \right)^3  years
\label{BHlfetime}
\end{equation}

\bigskip

\noindent
Because of observational constraints\cite{Alcock,Ostriker} 
the dark matter
constituents must generally be another twenty orders of magnitude more massive
than the lower limit in Eq.(\ref{PBHmin}).  We assert \cite{PF1,PF2,PF3}
that most dark
matter black holes are in the mass range between
one hundred and one hundred thousand times the solar mass.
The name intermediate mass black holes (IMBHs)
is appropriate because they lie in mass above stellar-mass black holes                                                                                                                                                                                             and below the
supermassive black holes which reside in galactic cores.

\bigskip

\noindent
The possibility that primordial black holes can be formed with intermediate
and higher masses has been established by an existence theorem
discussed in a related paper\cite{Frampton2}.

\bigskip

\noindent
Let us discuss three methods (there may be more)
which could
be used to search for dark matter IMBHs. 
While so doing we shall clarify 
what limits, if any, can be deduced from
present observational knowledge.

\bigskip

\noindent
Before proceeding, it is appropriate first
to mention the important Xu-Ostriker upper bound 
of about a million
solar masses from galactic disk stability\cite{OstrikerXu}
for any MACHO residing inside the galaxy.

\bigskip
\bigskip
\bigskip

\section{Wide Binaries}

\bigskip

\noindent
There exist in the Milky Way pairs of stars which are gravitationally bound binaries
with a separation more than 0.1pc. These wide binaries retain their original orbital parameters
unless compelled to change them by gravitational influences, for example, due to
nearby IMBHs.

\bigskip

\noindent
Because of their very low binding energy, wide binaries are particularly sensitive 
to gravitational perturbations and can be used
to place an upper limit on, or to detect, IMBHs. 
The history of employing this ingenious technique is regretfully checkered. 
In 2004 a fatally strong constraint was claimed by an Ohio State University group
\cite{Yoo} in a paper entitled
"End of the MACHO Era" so that, for researchers who have time
to read only titles and abstracts, stellar and higher mass constituents of
dark matter appeared to be totally excluded.

\bigskip

\noindent
Five years later in 2009, however, another group this time from Cambridge University
\cite{Quinn} reanalyzed the available data on wide binaries
and reached a quite different conclusion.
They questioned whether {\it any} rigorous constraint on MACHOs
could yet be claimed, especially as one of the important binaries
in the earlier sample had been misidentified.

\bigskip

\noindent
Because of its checkered history, it seems wisest to proceed with
caution but to recognize that wide binaries represent a potentially useful
source both of constraints on, and the possible discovery of, 
dark matter IMBHs.

\bigskip

\noindent
A further study of wide binaries \cite{Monroy} attempted to
place limits on MACHOs.  However, unlike microlensing which has 
positive signals, wide binary analysis is a null experiment and
we remain skeptical of any limits claimed. 

\section{Distortion of the CMB}

\bigskip

\noindent
This approach hinges on the phenomenon of accretion of gas onto the IMBHs.
The X-rays emitted by such accretion of gas are downgraded in frequency
by cosmic expansion and by Thomson scattering becoming microwaves which 
distort the CMB, both with regard to its spectrum and to its anisotropy.

\bigskip

\noindent
One impressive calculation of this effect \cite{Ostriker} employs a specific model for the
accretion, the Bondi-Hoyle model\cite{BondiHoyle}, and carries through the computation
all the way up to a point of comparison with data from FIRAS on CMB spectral distortions
\cite{FIRAS}, where FIRAS was a sensitive device attached to the COBE satellite.
Unfortunately the paper includes the limits from wide binaries discussed in \cite{Yoo},
{\it ut supra}, and preceded the corrective paper \cite{Quinn}, so its results might have
been influenced.

\bigskip

\noindent
The results obtained from this approach are interesting if one can be
certain that the gas accretion, subsequent X-ray emission and downgrading
are well modeled. Like wide
binaries, CMB distortion is indirect but could in future lead to useful bounds on,
or the possible discovery of, dark matter IMBHs.

\bigskip
\bigskip

\section{Microlensing}

\noindent
Microlensing is the most direct experimental method and has the big advantage
that it has successfully found
examples of MACHOs. The MACHO Collaboration used a method
which had been proposed\footnote{We have read that such gravitational lensing was later found to have been calculated 
in unpublished 1912 notes by Einstein who did not publish perhaps because at that time he considered its experimental measurement
impracticable.} by Paczynski\cite{Paczynski} where the amplification of a distant
source by an intermediate gravitational lens is observed. The MACHO Collaboration
discovered several striking microlensing events whose light curves are
exhibited in \cite{Alcock}. The method certainly worked well for $M < 100 M_{\odot}$
and so should work equally well for $M > 100 M_{\odot}$ provided one can devise
a suitable algorithm and computer program to scan enough sources.

\bigskip

\noindent
The longevity of a given lensing event is proportional to the square root of the lensing mass
and numerically is given by
($\hat{t}$ is longevity)
\begin{equation}
\hat{t} \simeq 0.2 yr \left( \frac{M_{lens}}{M_{\odot}} \right)^{1/2}
\label{that}
\end{equation}
where a transit velocity $200km/s$ is assumed for the lensing object.

\bigskip

\noindent
The MACHO Collaboration investigated lensing events with longevities
ranging between about two hours and two years. From Eq.(\ref{that}) this corresponds
to MACHO masses between approximately $10^{-6} M_{\odot}$ and $100 M_{\odot}$.

\bigskip

\noindent
The total number and masses of objects discovered by the MACHO Collaboration
could not account for all the dark matter known to exist in the Milky Way. At most
10\% could be explained. To our knowledge, the experiment ran out of money and
was essentially abandoned in about the year 2000.
But perhaps the MACHO Collaboration and its funding
agency were too easily discouraged.

\bigskip

\noindent
What is being suggested is that the other 90\% of the dark matter in the
Milky Way is in the form of MACHOs which are more massive than those detected
by the MACHO Collaboration, and which almost certainly could be detected by a
straightforward extension of their techniques. In particular, the expected
microlensing events have
a duration ranging up to two centuries.
Let us consider the entries of Table 1
which merit discussion both with respect to the proposed microlensing experiment and
briefly with respect to the entropy of the universe.

\bigskip

\noindent
We simplify the visible universe without losing anything important
by regarding it as containing exactly $10^{11}$ galaxies, each with mass
(dominantly dark matter) of exactly $10^{12} M_{\odot}$. The first three
columns of the Table consider one halo of dark matter. To
a first approximation, we can temporarily ignore the normal matter.
The fourth column gives the additive entropy of the universe
for well separated halos and the fifth column gives the corresponding microlensing event
longevity in years.

\bigskip

\noindent
For a black hole with mass $M_{BH} = \eta M_{\odot}$, the dimensionless
entropy is $S_{BH}/k  \sim 10^{77} \eta^2$, in other words
\begin{equation}
S_{BH} /k = 10^{77} \left( \frac{M_{BH}}{M_{\odot}} \right)^2.
\label{BHentropy}
\end{equation}

\bigskip

\noindent
If we study the first five rows of Table 1 we notice that, for a given total halo mass,
$M_{Halo} = 10^{12} M_{\odot}$, a smaller number of heavier black holes gives
higher entropy because $S_{BH} \propto M_{BH}^2$.
Within galactic halos, the black hole masses are restricted to be below
$10^6 M_{\odot}$ by the disk stability already mentioned.
Various arrangements of the allowed
black hole mass function were explored in \cite{Ludwick}.
Arguments using the concept of the entropy of the universe,
together with the second law of thermodynamics, are strongly suggestive
of many more black holes than the stellar and supermassive types
already identified for the simple reason that
black holes are, by far, the most efficient concentrators of entropy.

\begin{table}[htdp]
\caption{Microlensing Longevity ($\hat{t}$) for the case of 
$n$ IMBHs per halo. IMBH mass =$\eta M_{\odot}$. 
Halo mass = $10^{12} M_{\odot}$.
Universe mass = $10^{23} M_{\odot}$. See also \cite{PHFIMBH}.}
\begin{center}
\begin{tabular}{|c|c|c|c||c||}
\hline
$n$/Halo  & M = $\eta M_{\odot}$  & Halo Entropy & Universe Entropy & Longevity   \\
$Log_{10} n$ & $Log_{10}\eta$ & $Log_{10} (S_{Halo}/k)$ & $ Log_{10} (S_{Universe}/k) $ & $\hat{t}$ (years)\\
\hline
\hline
10 & 2 & 91 & 102 & 2  \\
\hline
9  & 3 & 92 & 103 &  6  \\
\hline
8  & 4 & 93 & 104 & 20  \\
\hline
7 & 5  & 94 & 105 & 60 \\
\hline
6 & 6  & 95 & 106 & 200 \\
\hline
\hline
{\it 0} & {\it 12} & {\it 101} & {\it 112} &  n/a \\
\hline
\hline
\end{tabular}
\end{center}
\label{longevity}
\end{table}

\bigskip

\noindent
The sixth
and last row in Table 1 illustrates how if a halo hypothetically collapsed into one
large black hole, its entropy would be
$S_{Halo}^{Max}/k  \sim 10^{101}$. If the superluminal accelerated
expansion prevents coalescence of such collapsed halos the additive entropy 
of the universe's interior would be $S_{Universe}/k \sim 10^{112}$.
If, hypothetically, all the halos would instead combine to one very large
black hole with mass
$10^{23} M_{\odot}$, the entropy would be $S_{Universe}/k \sim 10^{123}$.
The Schwarzschild radius of this very large black hole is $R = 10^{23} \times 3km
\sim 30 Gly$, not far below the comoving radius ($\sim 45Gly$) of the visible universe.

\bigskip

\noindent
The discussion of the previous paragraph implies, as has been discussed elsewhere, that the visible universe 
is, in some sense, close to itself being a black hole inside of which we live.
This curious fact seems to have no bearing on dark matter
but may be relevant to the more difficult problem of dark energy.

\bigskip

\noindent
Microlensing experiments involve systematic scans of millions of distant star
sources because it requires accurate alignment of the star and the
intermediate lensing MACHO. Because the experiments are already highly computer
intensive, it makes us more optimistic that the higher longevity events
can be successfully analyzed. Study of an event lasting two centuries should
not necessitate that long an amount of observation time.  It does require suitably ingenious 
computer programming to track light curves and distinguish them from
other variable sources. This experiment is undoubtedly extremely
challenging, but there seems no obvious reason it is impracticable.

\bigskip

\noindent
A fraction of the resources currently being thrown at WIMP searches
could be enough to support this desirable pursuit of high-longevity
microlensing observations.

\bigskip
\bigskip
\bigskip

\section{Discussion}

\noindent
Axions may not exist for theoretical reasons\cite{Holman,Kamionkowski,BarrSeckel}
discovered in 1992. Weak scale supersymmetry may not exist
for the experimental reason\cite{ATLASsusy,CMSsusy,CMS2} of its non-discovery 
at the LHC.
The idea that dark matter experiences weak interactions (WIMPs) came historically
from the appearance of an appealing DM constituent, the neutralino,
in the theory of weak scale supersymmetry.

\bigskip

\noindent
The only interaction which we know for certain to be experienced
by dark matter is gravity and the simplest assumption is that gravity
is the only force coupled to dark matter.
Why should the dark matter experience the weak interaction
when it does not experience the strong and electromagnetic interactions?
If it does not, then terrestrial experiments searching for dark matter by either direct
detection or production would be doomed to failure. 

\bigskip

\noindent
We began with four candidates for dark matter constituent:
(1) axion; (2) WIMP; (3) brown dwarf, understood to include all compact baryonic
objects;
(4) black hole.
We {\it pro tempore} eliminated the first two by hopefully persuasive
arguments, made within the context of an overview
of particle phenomenology. We eliminated the third by the upper limit
on baryons imposed by robust Big Bang 
Nucleosynthesis (BBN) calculations.

\bigskip

\noindent
We should remain open to the possibility that axions and /or WIMPs may exist and
that the dark matter might involve them as well, with PIMBHs giving only the
dominant contribution. We also point out that black holes as constituents
of dark matter in a general sense have been previously discussed {\it e.g.} in \cite{Linde} 
for many years.

\bigskip

\noindent
We assert that PIMBHs can constitute almost all  dark matter while maintaining
consistency with the BBN calculations. This is an important point because distinguished
astronomers have written an opposite assertion {\it e.g.} Begelman and Rees \cite{Rees}
state, on their page 256, that black holes cannot form more than 20 \% of dark matter
because the remainder is non-baryonic and Freese \cite{Freese} states similarly, on her page 99, that, because of the constraint imposed by BBN, astrophysical black holes cannot provide the dark matter in galaxies. 

\bigskip

\noindent
Both sources are making an implicit assumption which does not apply to 
the PIMBHs which we assert comprise almost all dark matter. That assumption is that
black holes can be formed only as the result of the gravitational collapse
of baryonic stars. We are claiming, on the contrary,  that dark matter black holes can be,
and the majority must be,
formed primordially in the early universe as calculated and demonstrated in \cite{Yanagida},
and endorsed by \cite{Carr}, unconstrained by the BBN upper limit on baryons. Earlier works
on PBHs, and DM candidates, include \cite{Kohri,Khlopov,Terazawa,Klimai}.

\bigskip

\noindent
Our proposal is that the Milky Way contains between ten million and ten billion
massive black holes each with between a hundred and a hundred thousand times the solar mass. Assuming the halo
is a sphere of radius a hundred thousand light years the typical separation
is between one hundred and one thousand light years which is also the most
probable distance of the nearest PIMBH to the Earth. At first sight, it may be
surprising that such a number of massive black holes could have
remained undetected in the Milky Way.
On second thoughts, it appears reasonable when one bears in mind their large mean
separation
of a hundred to a thousand light years and their relatively
small size, all being physically smaller than the Sun.

\bigskip

\noindent
Of the detection methods discussed, extended microlensing observations
seem the most promising and writing the present paper will have been
worthwhile if efforts to detect
higher longevity microlensing events are hereby encouraged.
It will be exceptionally rewarding if most of the dark matter in our galaxy is confirmed, by
microlensing techniques or otherwise, to be
in the form of intermediate-mass black holes.

\bigskip
\bigskip
\bigskip

\section*{Acknowledgement}

\noindent
We would like to acknowledge the late Professor David Cline of UCLA.

\end{document}